\documentclass[authoryear,preprint,review,12pt]{elsarticle}

\usepackage{amssymb}
\usepackage{xspace}
\usepackage{aas_macros}
\usepackage{textcmds}
\usepackage{newtxtext,newtxmath}

\usepackage[T1]{fontenc}
\usepackage{ae,aecompl}

\usepackage{graphicx}	
\usepackage{amsmath}	
\usepackage{changes}
\usepackage{relsize}

\journal{New Astronomy}

\def\apj{ApJ}
\def\apjl{ApJL}
\def\apjs{ApJ Suppl. Ser.}

\def\apspr{Astroph. Sp. Phys. Rev.}
\def\aap{A\&A}

\def\mnras{MNRAS}

\def\beq#1{\begin{equation}\label{#1}}
\def\eeq{\end{equation}}
\def\beqa#1{\begin{eqnarray}\label{#1}}
\def\eeqa{\end{eqnarray}}

\def\Eq#1{Eq.~(\ref{#1})} 
\def\eqn#1{~(\ref{#1})}
\def\myfrac#1#2{\left(\frac{#1}{#2}\right)}

\def\comment#1{\relax}

\newcommand{\pb}{P_\mathrm{b}}
\newcommand{\dpb}{\dot P_\mathrm{b}}
\newcommand{\mv}{M_\mathrm{V}}
\newcommand{\mx}{M_\mathrm{X}}
\newcommand{\dmv}{\dot M_\mathrm{V}}
\newcommand{\dmx}{\dot M_\mathrm{x}}

\begin{document}

\begin{frontmatter}


\title{Evolutionary Increase of the orbital Separation and Change of the Roche Lobe Size in SS433}

\author[sai]{Anatol Cherepashchuk}
\author[sai]{Alexander Belinski}
\author[sai]{Alexander Dodin}
\author[sai,kfu]{Konstantin Postnov}

\address[sai]{M.V. Lomonosov Moscow State University, Sternberg Astronomical Institute, 13, Universitetskij pr., 119234, Moscow, Russia}
\address[kfu]{Kazan Federal University, Kremlevskaya 18, 420008 Kazan, Russia}


\begin{abstract}
We present results of long-term photometric  monitoring of SS433 which proves a secular evolutionary increase of the orbital period of SS433 at a rate of $(1.14\pm 0.25)\times 10^{-7}$ s~s$^{-1}$. Using a physical model of non-conservative mass transfer in SS433 through a supercritical accretion disc around the compact companion, we reliably confirm that the binary mass ratio in SS433, $q=\mx/\mv$ is $\gtrsim 0.8$. For an optical star mass $\mv\sim 10 M_\odot$ the compact object in SS433 is a black hole with mass $M_{BH}\gtrsim 8 M_\odot$. We discuss  evolutionary implications of the found orbital period increase in SS433 -- a secular change in the orbital separation and a size of the Roche lobe of the optical star. We show that for the mass-loss rate $\dmv\sim 10^{-4}-3\times 10^{-5} M_\odot$
per year  and an optical star mass $\mv \sim 10-15 M_\odot$ the found orbital period increase implies the corresponding orbital separation increase while the Roche lobe size can shrink or expand around a mean constant value depending on the optical star mass-loss rate which may be modulated with the precessional period.
\end{abstract}

\end{frontmatter}



\section{Introduction}

The Galactic microquasar SS433 is an eclipsing massive X-ray binary at an advanced evolutionary stage with precessing supercritical accretion disc and relativistic jets \citep{1979ApJ...233L..63M,1979MNRAS.187P..13F,1979A&A....76L...3M,1980ApJ...235L.131C,1981MNRAS.194..761C,2020NewAR..8901542C}. This object has been intensively explored in the radio, IR, optical and X-rays (see, e.g., review by \cite{2004ASPRv..12....1F} and references therein). However, for more than 40 years of studies two important issues remained unresolved:
\begin{enumerate}
 \item What is the nature of the compact object in SS433 -- a neutron star or a black hole?
    \item Why does the massive binary system SS433 at the second mass-transfer stage evolve as a semi-detached system avoiding the common envelope formation? 
\end{enumerate}

Spectral studies and analysis of X-ray eclipses in SS433 yielded controversial results (see, for example, the discussion in \cite{2004ASPRv..12....1F}). This is largely due to a complexity of the physical model of SS433: a huge mass-loss rate from the optical star overfilling its Roche lobe $\dmv\sim 10^{-4}-10^{-5} M_\odot$ yr$^{-1}$, the presence of a rapidly rotating selectively absorbing circumbinary  shell, the accretion disc precession, etc.

\cite{2004ASPRv..12....1F} and \cite{2018MNRAS.479.4844C,2019MNRAS.485.2638C} proposed a different approach to estimate the basic parameters of SS433 using the astonishingly stable orbital period of this microquasar.  Later on, the analysis of our own long-term photometric observations of SS433 and data from the literature enabled us to obtain some evidence of a secular increase in the orbital period of SS433 suggesting a binary mass ratio $q=\mx/\mv\gtrsim 0.8$ \citep{2021MNRAS.507L..19C}. We concluded that the compact object in SS433 is a black hole. Therefore, the issue 1 can be considered as resolved. Moreover, the high binary mass ratio can provide a stable mass transfer rate from the optical star overfilling its Roche lobe onto a compact companion \citep{2017MNRAS.471.4256V}.

In the present paper, we report our new photometric observations of SS433 which confirm the secular increase of the orbital period of SS433 at a rate of $(1.14\pm 0.25)\times 10^{-7}$ s~s$^{-1}$ (Section 2). We also analyse implications of this result for the secular evolution of the orbital separation and Roche lobe size of the optical star in SS433 (Section 3). We find that in the frame of the adopted physical model of non-conservative mass transfer onto the compact object through a supercritical accretion disc with powerful wind and possible mass loss from the system via the outer Lagrangian point the orbital separation of SS433 should increase for the adopted binary system parameters and optical star mass-loss rate on a thermal time scale. The time derivative of the size of the optical star Roche lobe is found to be close to zero, possibly due to modulated mass-loss rate of the optical star with orbital and precessional periods. These findings help us to resolve issue 2 why the massive binary SS433 evolves as a semi-detached binary avoiding the common envelope stage. 

\section{Observations}

Observations of SS433 carried out in the period 1978-2020 were completed by new data (224 estimates in the $V$ band in 216 nights) obtained by the automatic telescope RC600 of the Caucasian Mountain Observatory (CMO SAI) in 2021-2022. The details of observations and data reduction are the same as in our previous paper \citep{2021MNRAS.507L..19C} and are described in \cite{2022ARep...66..451C}.  

To calculate the phase delays in the orbital variability of SS433 we reproduced the steps described in \cite{2021MNRAS.507L..19C} with the same template $\overline{V}(\varphi)$ for the orbital light curve. As before, we consider time intervals within the precession phase interval  $T_3\pm0.2P_{\rm prec}$ corresponding to the maximum disc opening to the observer, when the orbital light curve of the source exhibits the most regular shape. 81 out of 224 new measurements in 2021-2022  meet this condition.

\begin{figure}
	\includegraphics[width=\columnwidth]{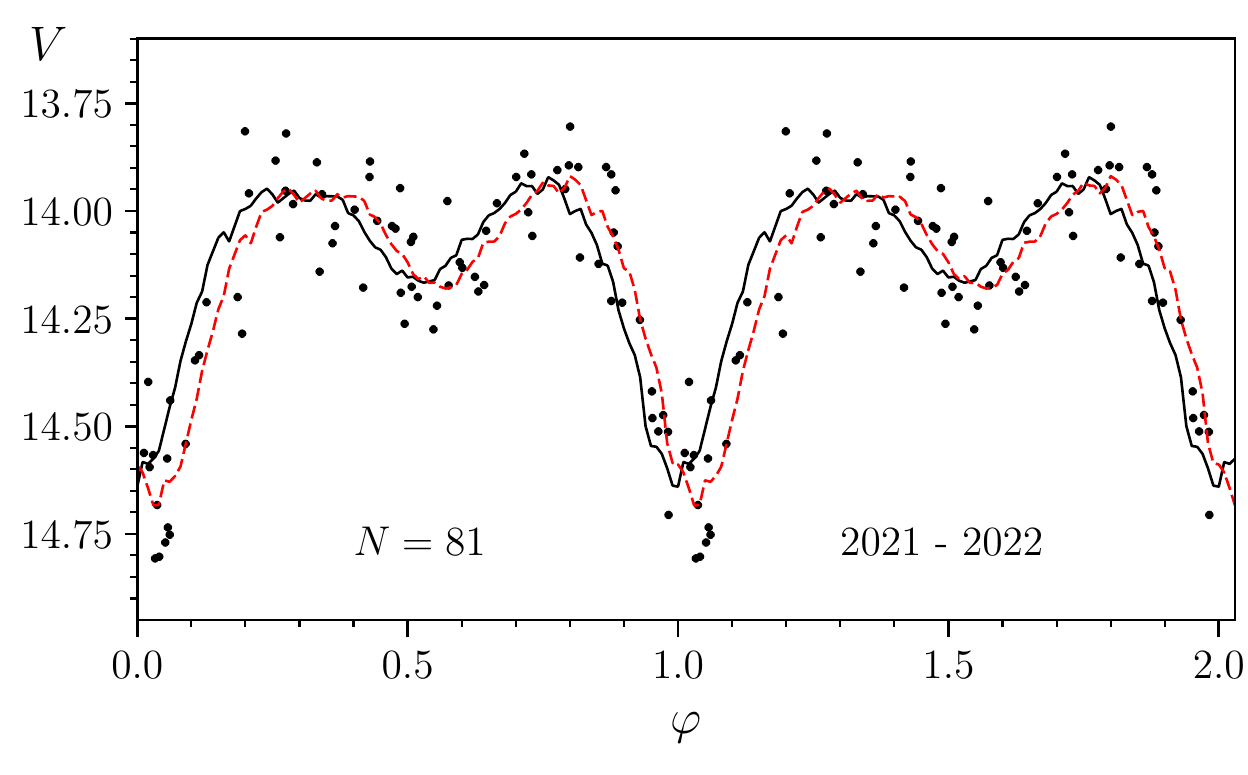}
    \caption{Observed light curve of SS433 within $T_3\pm0.2P_{\rm prec}$ phased with the orbital period in 2021-2022. The solid black line is the orbital light curve template. The red dashed lines shows the scaled and shifted light curve best-fitting the observations using Method 1 (Hertzsprung's method).
    }
    \label{f:orbcurve}
\end{figure}

Like in our earlier papers we have used two methods to derive the phase delays:

\textit{Method 1}. Inside each time interval, the template $\overline{V}(\varphi)$ was fitted to the photometric data $V(\varphi)$ around $T_3\pm0.2P_{\rm prec}$ by adjusting three free parameters $a$, $\Delta \varphi$, $c:$
$V(\varphi) = a\overline V(\varphi-\Delta \varphi)+c.$
This is the well known Hertzsprung method.
See Fig.\,\ref{f:orbcurve} for the last time interval 2021-2022.

\textit{Method 2}. The primary minimum inside the orbital phase interval $-0.15<\varphi<0.15$ is approximated by a Gaussian with four free parameters.

\begin{figure}
	\includegraphics[width=\columnwidth]{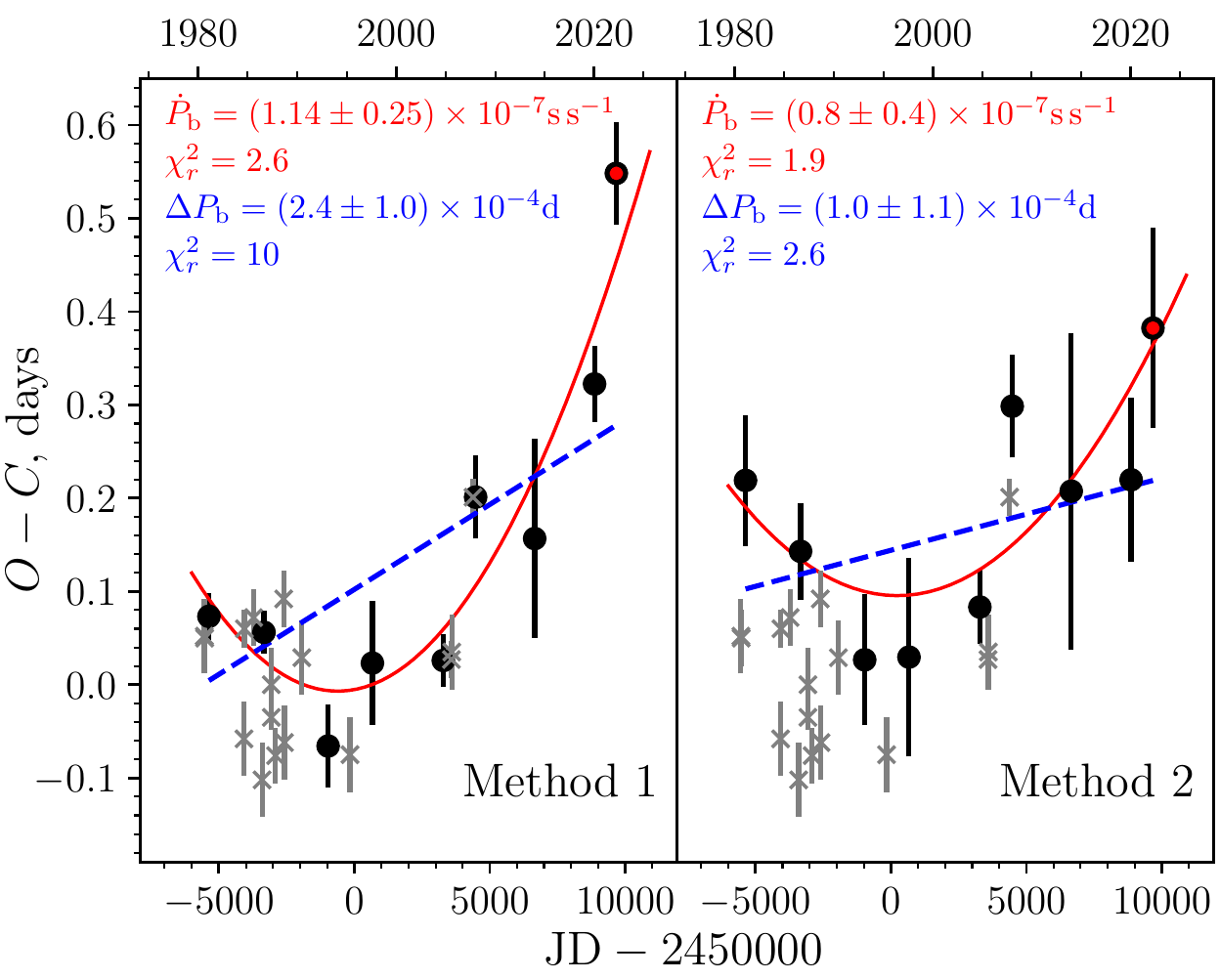}
    \caption{$O-C$ residuals of SS433 relative to the ephemeris with a constant orbital period of $P_\mathrm{b}=13^\mathrm{d}.08223$ calculated by different methods. The red filled circle marks the new value of $O-C$. Grey crosses are $O-C$ residuals for data from the literature. Red solid curves show a parabolic fit, blue dashed lines correspond to a linear fit. 
    The orbital period derivative $\dot{P}_\mathrm{b}$     
    and the correction to the constant orbital period $\Delta P_\mathrm{b}$ are shown for each case.
    }
    \label{f:oc}
\end{figure}

The obtained phase delay is converted to the time delay $(O-C)$ using the mean orbital period $P_{\rm b}.$
In Fig. \ref{f:oc} we add the new values $(O-C)_{1,2}$ to Figure 2 from \cite{2021MNRAS.507L..19C} and recalculated the coefficients. New data fully confirm the previously suspected increase in the orbital period with $\dpb=(1.14\pm0.25)\times 10^{-7}$\,s\,s$^{-1}$ (the previous estimate was $\dpb=(1.0\pm0.3)\times 10^{-7}$\,s\,s$^{-1}$). Method 2 confirms the period increase but with a poorer accuracy. The new observations allow us to more reliably reject the third-body hypothesis, since a mass of at least $20 {\rm M}_{\odot}$ is now required to explain the observed $(O-C)$ by the third body.

To summarize, the observed primary minima of SS433 and the refined ephemeris are\\
\begin{tabular}{ccc}
2444633.94 $\pm$ 0.03,&2446661.67 $\pm$ 0.02,&2449029.43 $\pm$ 0.04,\\
2450664.80 $\pm$ 0.07,&2453281.25 $\pm$ 0.03,&2454471.91 $\pm$ 0.04,\\
2456656.59 $\pm$ 0.11,&2458867.66 $\pm$ 0.04,&2459665.90 $\pm$ 0.06;\\
 \end{tabular}
 \begin{center}
\begin{tabular}{crrl}
 $T_{\rm min}=$ &2451737.54 &  $+13.08250E$       &$+7.3\times10^{-7}E^2.$ \\
               &  $\pm$.03 &   $\pm$.00005\,~\,\,&  $\pm$1.6 
\end{tabular}
\end{center}    

\section{Implications of the secular increase of $\pb$ to the orbital separation and Roche lobe size changes.}

We consider a physical model with complicated mass flows in SS433 using the following assumptions:
\begin{itemize}
    \item a non-conservative mass exchange occurs from the optical star $\mv$  onto the compact object $\mx$;
    \item the total mass-loss from the binary system is fully determined by the mass-loss from the optical star $\dmv$, i.e. we will neglect the compact star mass growth, $\dmx=0$;
    \item isotropic re-emission of the fraction $\beta$ of mass transferred from the optical star occurs via the supercritical accretion disc wind;
    \item the fraction of the mass-loss rate $(1-\beta)\dmv$ occurs through the outer Lagrangian point $L_2$ via a circumbinary disc with specific angular momentum (in units of the orbital angular momentum) $K$
\end{itemize}
(see \cite{2018MNRAS.479.4844C, 2019MNRAS.485.2638C,2021MNRAS.507L..19C} for more detail and derivations). 

In this model, the binary orbital period fractional change reads 
\beq{e:dPP}
\frac{\dot P_\mathrm{b}}{P_\mathrm{b}}=-\myfrac{\dot M_\mathrm{V}}{M_\mathrm{V}}\myfrac{3q^2+2q-3\beta -3K(1-\beta)(1+q)^{5/3}}{q(1+q)}\,.
\eeq
We have used this equation in \citep{2021MNRAS.507L..19C} to constrain the binary system mass ratio $q=\mx/\mv$ based on the measured value of the fractional orbital period change $\dpb/\pb$, the specific angular momentum  of matter escaping from the system via the circumbinary disc $K$, the assumed value of $\beta\le 1$ and the fractional mass-loss rate from the optical star $\dmv/\mv$. Our analysis led to the estimate $q\gtrsim 0.8$, $\mx\gtrsim 8 M_\odot$ \citep{2021MNRAS.507L..19C}. 

For possible evolutionary implications of the measured increase of the binary orbital period, it is instructive to consider the change in the orbital separation. Should it increase in response to the mass transfer, the Roche lobe of the donor star will shrink less rapidly (or even expand, see below) helping to avoid a possible instability leading to the formation of a common envelope. 

\begin{figure*}
	\includegraphics[width=\textwidth]{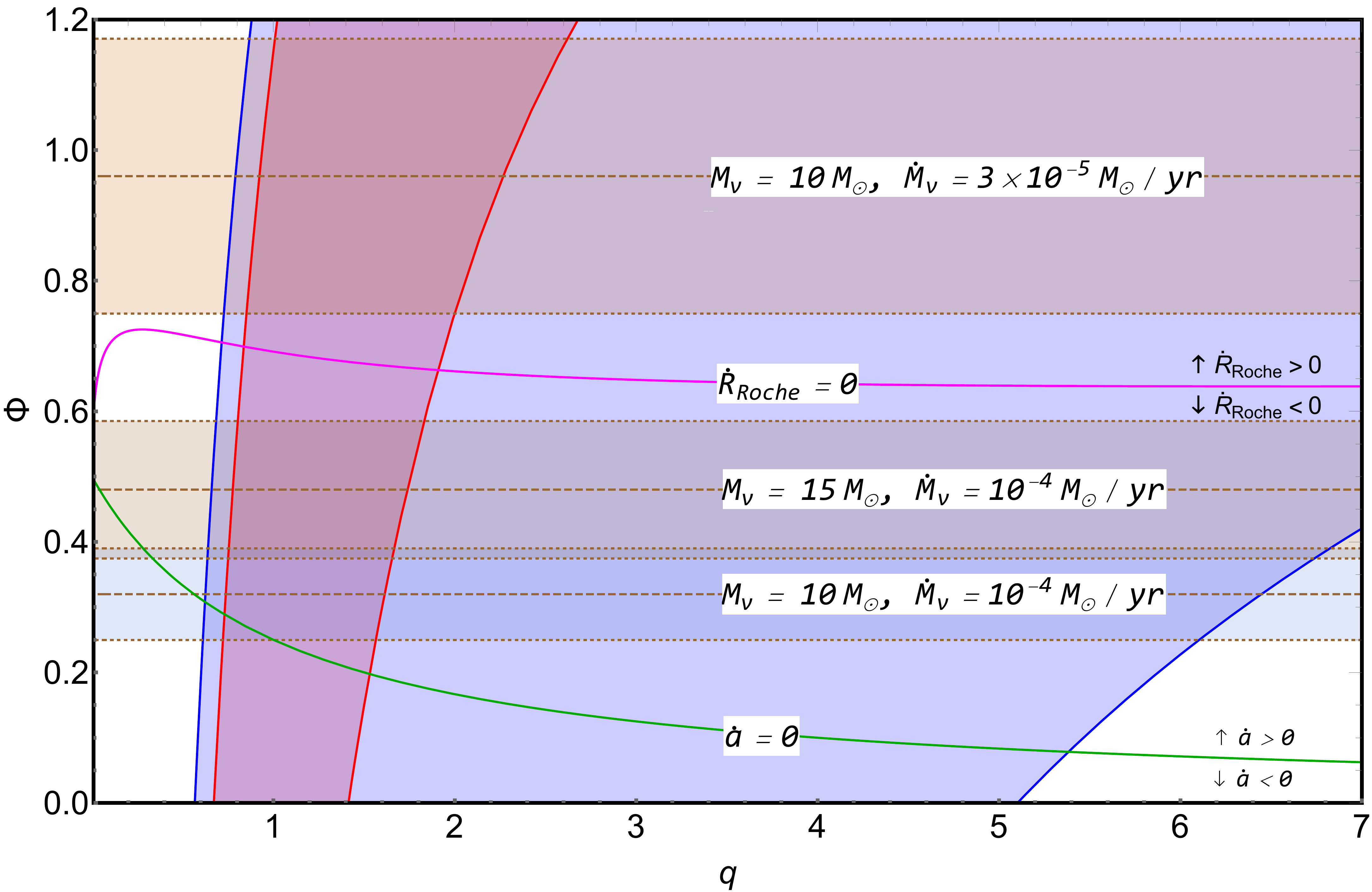}
    \caption{The dimensionless fractional orbital period change in SS433 normalized to the fractional mass-loss rate from the optical star, $\Phi(q,\beta,K)=(\dpb/\pb)/(|\dmv|/\mv)$, as a function of the binary mass ratio $q$ [\Eq{e:Phi}]. 
    The narrow shadowed area between the solid almost vertical red lines corresponds to the case $\beta=0.9$ (10\% of the mass loss through the outer Lagrangian point) with the specific angular momentum in a circumbinary disk $K=0$ (the left boundary) and $K=4.7$ (the right boundary). The wide shadowed area corresponds to the case $\beta=0.7$ (30\% of the mass loss through the outer Lagrangian point). The green hyperbola corresponds to zero time derivative of the binary separation ($\dot a=0$, \Eq{e:daadmm}). The binary separation increases above this hyperbola. The magenta curve shows zero time derivative of the optical star Roche lobe corresponding to equality in \Eq{e:dRV+}. The Roche lobe of the optical star increases above this curve.
    Between the magenta and green curves the orbital separation increases but the Roche lobe size of a binary with mass ratio $q$ decreases. The hatched horizontal strips bounded by brown dotted lines correspond (from top to bottom) to the observed $\dpb=(1.14\pm 0.25)\times 10^{-7}$ s~s$^{-1}$ for three cases ($\mv=10 M_\odot$, $\dmv=3\times 10^{-5} M_\odot$ per year), ($\mv=15 M_\odot$ , $\dmv=10^{-4} M_\odot$ per year) and  ($\mv=10 M_\odot$, $\dmv=10^{-4} M_\odot$ per year).  }
    \label{f:Phi(q)}
\end{figure*}

As $\dmv$ is always negative, the sign of $\dpb$ is fully determined by the parameters $(q,\beta, K)$, so it is convenient to introduce the dimensionless quantity 
\beq{e:Phi}
\Phi(q,\beta,K)=\frac{\dpb/\pb}{|\dmv|/\mv}=\frac{d\ln\pb}{|d\ln\mv|}
\eeq
(denoted as $-A$ in our previous paper \cite{2021MNRAS.507L..19C}). Using the 3rd Kepler's law, it is straightforward to write down the fractional change rate of the  binary orbital separation $\dot a/a$:
\beq{e:daa}
\frac{\dot a}{a}=\frac{2}{3}\myfrac{\dpb}{\pb}+\frac{1}{3}\myfrac{\dmv+\dmx}{\mv+\mx}=\frac{2}{3}\myfrac{\dpb}{\pb}-\frac{1}{3}\myfrac{|\dmv|}{\mv}\myfrac{1}{1+q}\,.
\eeq
The fractional change rate of the orbital separation in units of the (positive) fractional change rate of the optical star mass $|\dmv|/\mv$ reads
\beq{e:daadmm}
\frac{\dot a/a}{|\dmv|/\mv}= \frac{2}{3}\Phi(q,\beta,K)-\frac{1}{3}\frac{1}{1+q}\,.
\eeq
The orbital separation $a$ will increase (positive $\dot a$) if
\beq{e:da+}
\Phi(q,\beta,K)>\frac{1}{2}\frac{1}{1+q}\,.
\eeq
Thus, the positive  change in the orbital period of SS433 derived from observations, $\dpb=(1.14\pm 0.25)\times 10^{-7}$ s~s$^{-1}$, together with the condition (\ref{e:da+}), can constrain the binary mass ratio $q$ in SS433 such that the observed increase in the orbital period be simultaneously accompanied by the \textit{increase} in the orbital separation. Clearly, in this case the system does not evolve to a common envelope irrespective of details of the mass loss from the system. Figure \ref{f:Phi(q)} illustrates these constraints.

In Fig. \ref{f:Phi(q)} we plot the dimensionless quantity $\Phi(q,\beta,K)$ \eqn{e:Phi} as a function of the binary mass ratio $q$ for different parameters $\beta=[1,0]$ (the fraction of the mass-loss via the isotropic Jeans outflow from the supercritical accretion disc around the compact star) and specific angular momentum $K$ carried away via a circumbinary disc, in the range $K=[0,4.7]$ (the upper limit was derived by us in \cite{2021MNRAS.507L..19C} from GRAVITY observations of SS433 \citep{2019A&A...623A..47W}). The shaded vertical areas 
correspond to the case of $\beta=0.7$ and 0.9 (the broad strip limited by blue lines and narrow strip limited by red lines, respectively) with parameter $K$ varying from 0 (left boundary) to 4.7 (right boundary). The green hyperbola corresponds to zero orbital separation time derivative ($\dot a=0$) from \Eq{e:daadmm}. Above this hyperbola, $\dot a>0$. Three hatched horizontal strips show the limits for the observed $\dpb=(1.14\pm 0.25)\times 10^{-7}$ s~s$^{-1}$ for three cases ($\mv=10 M_\odot$, $\dmv=3\times 10^{-5} M_\odot$ per year), ($\mv=15 M_\odot$ , $\dmv=10^{-4} M_\odot$ per year) and  ($\mv=10 M_\odot$, $\dmv=10^{-4} M_\odot$ per year)
(the upper, middle and bottom strips, respectively). It is seen that for $q\gtrsim 0.8$ in all three cases the orbital period and orbital separation increase simultaneously. 

From the point of view of the binary evolution, it is also interesting to understand how the Roche lobe of the optical donor star responds to the mass loss. The Roche lobe of the optical star is determined by the orbital separation $a$ and the mass ratio $q$: $R_\mathrm{V}=af_\mathrm{V}(q)$, where $f(q)$ can be approximated according to \cite{1983ApJ...268..368E} as
\beq{e:RV}
f_\mathrm{V}(q)=\frac{0.49}{0.6+q^{2/3}\ln(1+q^{-1/3})}
\eeq
(this equation is derived from Eggleton's formula by substituting $q\to 1/q$
according to our choice of the mass ratio $q=\mx/\mv$). Then the fractional change rate of the donor's Roche lobe is
\beq{}
\frac{\dot R_\mathrm{V}}{ R_\mathrm{V}}= \frac{\dot a}{a}+\myfrac{d\ln f_\mathrm{V}}{d \ln q}\myfrac{\dot q}{q}=\frac{\dot a}{a}+\myfrac{|\dot \mv|}{\mv}\myfrac{d\ln f_\mathrm{V}}{d \ln q}\,.
\eeq
(In the last equality we took into account that in our case where $\dot \mx=0$, $\dot q/q=-\dot \mv/\mv=+|\dot \mv|/\mv$). The logarithmic derivative of $f_\mathrm{V}(q)$ reads
\beq{e:dfq}
\frac{d\ln f_\mathrm{V}}{d \ln q}=  -
\frac{2}{3}\left\{1-\left[
\frac{0.6}{0.49}+
\frac{1}{2}\frac{q^{2/3}}{1+q^{1/3}}\right]f_\mathrm{V}\right\}\,.
\eeq
Factorizing out the positive coefficient $|\dot \mv|/\mv$ as above and making use of \Eq{e:daa}, we obtain the condition for the fractional size of the donor's Roche lobe not to decrease, $\dot R_\mathrm{V}/R_\mathrm{V}\ge 0$, in the form
\beq{e:dRV+}
\Phi(q,\beta,K)\ge \frac{1}{2}\frac{1}{1+q}+
\left\{1-\left[
\frac{0.6}{0.49}+
\frac{1}{2}\frac{q^{2/3}}{1+q^{1/3}}\right]f_\mathrm{V}\right\}\,.
\eeq
It is easy to verify that the right-hand side of this inequality reaches a minimal value of 0.5 at $q\to 0$, takes a maximum value of $\approx 0.73$ at $q\approx 0.28$ and stays almost constant $\sim 0.65$ at $q> 1$ (see the magenta curve in Fig. \ref{f:Phi(q)}). Therefore, for the Roche lobe of the optical star to increase in SS433 with the observed value of the orbital period derivative, the mass-loss rate of the optical star should satisfy the inequality $\dot \mv<6\times 10^{-5}[M_\odot \mathrm{yr}^{-1}](\mv/10 M_\odot)$, which is likely for an optical star $\mv=10M_\odot$ losing mass in a thermal time scale $\dmv\sim 3\times 10^{-5}[M_\odot \mathrm{yr}^{-1}]$. 

\section{Discussion}
\label{s:discussion}

The increase in the orbital period of SS433, confirmed by our new photometric observations,  enabled us to prove that the binary mass ratio is $q\gtrsim 0.8$. As the mass of the optical star in SS433 has been independently inferred by the photometric method using the distance to SS433 $d=5.5$~kpc to be $\mv\approx 10 M_\odot$ \citep{2011PZ.....31....5G}, the mass of the compact object should be $\mx\gtrsim 8 M_\odot$, reliably classifying it as a black hole. In our model of SS433, the orbital period change occurs for several counteracting reasons: 
\begin{enumerate}
    \item The mass transfer from the optical star through the inner Lagrangian point $L_1$, which in the conservative case would decrease the orbital separation.
    \item The non-conservative mass-loss from the system via the outer Lagrangian point $L_2$  decreasing the orbital angular momentum and correspondingly the orbital separation.
    \item The radial wind outflow from a supercritical accretion disc (the Jeans isotropic re-emission) leading to orbital separation increase.
\end{enumerate}
 The balance between these outflows depends on the binary mass ratio $q$, the assumed fraction of non-conservative mass transfer $\beta$ and the specific angular momentum carried away via a circumbinary disc $K$ (see \Eq{e:dPP}). For $q\gtrsim 0.8$ the orbital period increases at the measured rate of $\dpb \approx 1.14\times 10^{-7}$ s s$^{-1}$. A low mass ratio $q<0.2$ (corresponding to the neutron star mass) would lead to the orbital period \textit{decrease} for any values of the parameters $\beta$ and $K$. We stress that our estimate of $q$ is obtained from independent dynamical considerations without  relying upon radial velocity curves and the analysis of X-ray eclipses in SS433.

 We stress that for an optical star mass $\mv=10-15 M_\odot$ and mass-loss rate $\dmv=10^{-4}-3\times 10^{-5} M_\odot$ per year, as seen in Fig. \ref{f:Phi(q)}, $q\sim 1$ implies that both the orbital period and orbital separation in SS433 increase simultaneously.
 
 The orbital separation increase explains the unusual, at first glance, fact that SS433 evolves as a semi-detached binary avoiding common envelope. \cite{2017MNRAS.471.4256V} showed that when a massive donor star in a close binary system fills or overfills its Roche lobe transferring mass onto a compact companion star, the binary system may avoid the formation of common envelope provided that the mass ratio of the components $q$ is sufficiently high. Such a system evolves as a semi-detached binary with a stable mass transfer onto the compact companion with the likely formation of a supercritical accretion disc. The mass and angular momentum outflow from the system occurs via a powerful wind from the innermost parts of the supercritical accretion disc (the so-called isotropic mass re-emission, or SS433-like mode). If the mass ratio is low, $q\lesssim 0.29$, a high-mass X-ray binary necessarily evolves to the common envelope stage. Our dynamical estimate of the binary mass ratio $q\gtrsim 0.8$ is in agreement with these predictions.

We have also analysed the fractional change of the Roche lobe size of the optical star in response to the mass transfer in SS433. If the  optical star mass is $\mv=10-15 M_\odot$ and the mass-loss rate $\dmv=10^{-4} M_\odot$ per year, the Roche lobe size decreases (the two bottom strips below the magenta curve in Fig. \ref{f:Phi(q)}). 
In this case, we should expect the formation of a common envelope because the thermally equilibrium radius of the optical component in SS433 (an A7I supergiant at an advanced stage of the nuclear burning)  should increase during the mass loss \citep{1994inbi.conf.....S}.  

For a 10 $M_\odot$ main-sequence star, the mass-loss rate this high corresponds to the thermal time scale $\tau_{KH}=GM^2/RL\propto M^{-2}$ so that $\dmv\sim \mv/\tau_{KH}\approx 3\times 10^{-5}[M_\odot \mathrm{yr}^{-1}] (\mv/10 M_\odot)^3$. 
For a 10 $M_\odot$ star with mass-loss rate  on a thermal time scale $\dmv\sim 3\times 10^{-5} M_\odot$ per year, the observed value of the orbital period increase implies that the Roche lobe size also increases (the upper strip above the magenta curve in Fig. \ref{f:Phi(q)}). The observed high mass-loss rate on a thermal time scale of the optical star sustained in SS433 suggests that the radius of the donor star keeps pace with the Roche lobe size. Therefore, the most plausible is 
an almost constant or increasing Roche lobe (the region above the magenta curve in Fig. \ref{f:Phi(q)}) involving a $\sim 10 M_\odot$ A7I donor star loosing mass on a thermal time scale. 


The sensitivity of the sign of the Roche lobe size derivative to the fractional mass loss rate $\dmv/\mv$ also may suggest that its variation 
during the precessional period of SS433 could lead to a specific self-regulation of mass-loss transfer, because the optical star's axial precession can change conditions for the mass transfer through the vicinity of the inner Lagrangian point. Therefore, it would be interesting to search for traces of variability in the mass-loss rate from the accretion disc in SS433 with the precessional period from spectroscopic observations. 
Indeed, \cite{2022ARep...66..451C} have found that the flux on the stationary H$_\alpha$ line emission in SS433 varies with the precessional phase by a factor of two reaching maximum at the phase of the maximum open disc to the observer and minimum at the disc-edge phase. Although this effect can be partially explained by variable projection of the disc on the sky (which is less probable because of the large H$_\alpha$ emitting volume), it is possible to assume that it also partially reflects the variability of $\dmv$ due to the precessing optical component.

Note that \citet{2007A&A...474..903B}, \citet{2008ARep...52..487D} and \citet{2022ARep...66..451C} found a quasi-periodicity in the velocity variations of the relativistic jets in SS433 with the orbital period and suggested that these variations can be caused by orbital ellipticity. Our discovery of the orbital ellipticity of SS433 \citep{2021MNRAS.507L..19C} enabled us to associate these variations with changing mass inflow into the accretion disc modulated by the motion of the optical star in the eccentric orbit. The optical star axial precession in the model of the slaved accretion disc in SS433 can also modulate the mass inflow into the supercritical accretion disc. 
 
\section{Conclusions}

Our new photometric observations of SS433 carried out in 2021-2022 confirmed the secular orbital increase of SS433 at a rate $\dpb=(1.14\pm 0.25)\times 10^{-7}$ s s$^{-1}$. Assuming a physical model of SS433 based on extensive multiwavelength observations, which includes a non-conservative mass transfer onto the compact object with isotropic re-emission from a supercritical accretion disc and mass outflow via the outer Lagrangian point $L_2$, we were able to derive a reliable estimate of the binary mass ratio in SS433 $q\gtrsim 0.8$ \citep{2021MNRAS.507L..19C}. The obtained lower limit on $q$ suggests the mass of the compact object in SS433 $\mx\gtrsim 8 M_\odot$, i.e. the compact object in SS433 is a black hole with the mass typical for Galactic black-hole X-ray binaries (e.g. \cite{2016ApJS..222...15T}).

In our model, a low binary mass ratio (corresponding to a neutron star as compact object) would lead to a secular \textit{decrease} in the orbital period of SS433, in contrast to observations. Therefore, the presence of a neutron star in SS433 should be rejected. 

We have shown that with an optical star mass $\mv=10-15 M_\odot$ and mass-loss rate  $\dmv\approx 3\times 10^{-5}-10^{-4} M_\odot$ yr$^{-1}$, which is close to the the expected value on a thermal time scale, the found value of the fractional orbital period increase in SS433 implies an \emph{increase} in the orbital separation of SS433. We have also explored the fractional change on the size of the Roche lobe of the optical star in SS443. Interestingly, the parameters of SS433 are close to the situation where the Roche lobe size stays almost constant during the mass transfer, which can explain why the system evolves as a semi-detached binary and avoid forming a common envelope.

Further evolution of this high-mass X-ray binary may lead to the formation of a neutron star -- black hole binary similar to GW200105 \citep{2021ApJ...915L...5A}. 

\section*{Acknowledgements}

We thank the anonymous referee for a careful reading of the paper and useful comments. 
We thank Dr. N. Ikonnikova and M. Burlak for help in RC600 observations. 
KP acknowledges the International Space Science Institute (ISSI) in Bern, through ISSI International Team project 512 Multiwavelength View on Massive Stars in the Era of Multimessenger Astronomy. AMC, AVD and AAB acknowledge the RSF grant 23-12-00092.
The work is supported by the Scientific and Educational School of M.V.  Lomonosov Moscow State University
'Fundamental and applied space research'. The observations are supported from M.V. Lomonosov Moscow State University Program of Development.

\section*{Data availability}
All CMO RC600 photometric observations of SS433 are  available on reasonable request from the authors.






\end{document}